\documentclass[conference]{IEEEtran}
\IEEEoverridecommandlockouts

\usepackage{amsmath,amssymb}
\usepackage{booktabs}
\usepackage{graphicx}
\usepackage{multirow}
\usepackage{array}
\usepackage{url}
\usepackage[hidelinks]{hyperref}
\usepackage{cite}
\usepackage{xcolor}
\usepackage{tikz}
\usetikzlibrary{arrows.meta,positioning,fit}
\usepackage{balance}
\usepackage{microtype}

\newcommand{\cvb}{CV-QRC$_b$}
\newcommand{\cvh}{CV-QRC$_h$}
\newcommand{\gbq}{GB-QRC}
\newcommand{\gco}{gCO$_2$e}

\title{Sustained Performance and Energy Accounting for Nonlinear Forecasting Across Classical and Simulated Quantum Models}

\author{
\IEEEauthorblockN{Avyay Kodali}
\IEEEauthorblockA{
Walnut Grove High School\\
Dallas, United States\\
avyaykodali03@gmail.com}
\and
\IEEEauthorblockN{Priyanshi Singh}
\IEEEauthorblockA{
Department of Computer Science and Engineering\\
SRM Institute of Science and Technology\\
Chennai, India\\
priyanshisingh.10009@gmail.com}
\and
\IEEEauthorblockN{Pranay Pandey}
\IEEEauthorblockA{
QuantumAI Lab, Fractal AI Research\\
Gurugram, India\\
pranay.pandey@fractal.ai}
\and
\IEEEauthorblockN{Krishna Bhatia}
\IEEEauthorblockA{
QuantumAI Lab, Fractal AI Research\\
Mumbai, India\\
krishna.bhatia@fractal.ai}
\and
\IEEEauthorblockN{Shalini Devendrababu}
\IEEEauthorblockA{
QuantumAI Lab, Fractal AI Research\\
Gurugram, India\\
shalini.devendrababu@fractal.ai}
\and
\IEEEauthorblockN{Srinjoy Ganguly}
\IEEEauthorblockA{
Department of Physics and Astronomy\\
University College London\\
London, United Kingdom\\
srinjoyganguly@gmail.com}
}

\begin{document}
\maketitle

\begin{abstract}
Energy-efficient AI requires more than reporting the lowest error or the shortest training time. The complete application path---data preparation, feature generation, model fitting, inference, memory use, and reusable intermediate state---determines sustained performance. We study this issue through nonlinear time-series forecasting, using simulated quantum reservoir computing (QRC) as an emerging-computing case study. The evaluation covers 33 forecasting configurations and 825 completed runs on NARMA-10, NARMA-20, Mackey--Glass, Lorenz-63, and Santa Fe laser data. The core artifact contains 775 fully instrumented runs across statistical, linear, reservoir, neural, continuous-variable Gaussian QRC, and gate-based statevector QRC models; 50 additional variational QNN runs extend the trainable-quantum comparison.

We record NRMSE, RMSE, MAE, wall time, inference latency, peak CPU/GPU memory, parameter count, operational energy, and carbon. Because a fixed QRC encoder can generate features once and serve several readouts, we report both cold-start and amortized accounting rather than hiding or repeatedly charging the same cache cost. We also use a Sustainable Forecasting Score (SFS) as a diagnostic geometric mean of normalized skill and log-scaled carbon efficiency, while retaining all raw measurements and Pareto views. The best average NRMSE is obtained by a baseline continuous-variable QRC with a Transformer readout (0.332), closely followed by classical TCN (0.341) and QRC+TCN (0.340). However, ESN and ridge-lag provide the strongest sustained efficiency, with average amortized carbon of 0.014 and 0.012 \gco{} per run and SFS values of 0.814 and 0.805. Enlarging the QRC feature map increases cost without improving mean accuracy, and gate-based statevector simulation is not competitive. The results motivate three systems practices for emerging AI workloads: expose stage-level energy, report reuse-aware accounting boundaries, and treat the accelerator/feature-map and classical readout as a co-designed application rather than isolated components.
\end{abstract}

\begin{IEEEkeywords}
energy-efficient AI, sustainable performance, time-series forecasting, quantum reservoir computing, energy measurement, carbon accounting, feature caching, system--application co-design
\end{IEEEkeywords}

\section{Introduction}
Peak accuracy is an incomplete systems metric. A forecasting workload may be accurate but expensive to train, slow to serve, memory intensive, or dependent on an accelerator whose preprocessing and feature-generation cost is omitted. Conversely, a model may consume little energy because it fails to extract useful signal. Energy-efficient computing therefore requires sustained application performance: useful work delivered under explicit time, power, memory, and carbon constraints. This perspective has long motivated performance-per-watt rankings in high-performance computing (HPC) \cite{feng2007green500} and is increasingly important for AI workloads \cite{schwartz2020green,henderson2020systematic,lannelongue2021green,dodge2022carbon}.

Nonlinear time-series forecasting is a compact but demanding setting in which to study these trade-offs. Standard benchmarks probe long memory, multiscale dynamics, chaotic sensitivity, and nonstationarity. They also require repeated executions across datasets, seeds, architectures, and readouts. The accumulated cost of a benchmark can therefore be much larger than a single headline training run. This is especially relevant when evaluating emerging computing technologies on classical simulators. Simulator choice, state dimension, circuit evaluations, feature materialization, host-device transfers, and downstream learning can dominate the measured cost even when the proposed algorithm trains few parameters.

We use quantum reservoir computing (QRC) as a case study. QRC applies fixed quantum or quantum-inspired dynamics to a temporal input and trains only a classical readout \cite{fujii2017harnessing,nakajima2019boosting}. This removes variational optimization of the reservoir and makes QRC attractive for near-term temporal learning. It also creates a systems-level accounting problem: reservoir features may be expensive to generate but reusable across linear, recurrent, convolutional, or attention-based readouts. Ignoring feature generation overstates efficiency; charging the entire feature cost to every readout understates the value of reuse. Further, a simulator-based result measures the host-side execution path, not future QPU energy.

This paper evaluates nonlinear forecasting through the lens of energy efficiency and sustained application performance. Our research questions are:
\begin{itemize}
    \item \textbf{RQ1:} Which model families deliver the best forecasting accuracy, and at what operational cost?
    \item \textbf{RQ2:} How do runtime, memory, energy, carbon, and accuracy jointly change the ranking?
    \item \textbf{RQ3:} How strongly do cache reuse and accounting boundaries affect QRC conclusions?
    \item \textbf{RQ4:} What system--application co-design lessons transfer to other accelerator-backed AI workloads?
\end{itemize}

The contributions are fourfold. First, we provide a common execution and reporting protocol for 33 classical, neural, variational-quantum, and QRC configurations over five nonlinear forecasting tasks. Second, the benchmark records accuracy, stage-level runtime, latency, memory, energy, and carbon for each dataset--model--seed run. Third, we formalize cold-start and amortized accounting for reusable feature maps. Fourth, we derive operational recommendations for sustainable benchmarking of emerging compute: measure the complete path, preserve raw metrics alongside composites, and schedule reusable feature generation as a shared service rather than an invisible preprocessing step.

\section{Background and Related Work}
\subsection{Energy-Efficient AI and HPC Measurement}
Green AI argues that computational efficiency should be reported as a first-class result rather than treated as an implementation detail \cite{schwartz2020green}. Carbon accounting studies show that emissions depend on energy, hardware utilization, location, grid intensity, and datacenter overhead \cite{strubell2019energy,lacoste2019quantifying,patterson2021carbon,dodge2022carbon}. Tools such as CarbonTracker and CodeCarbon make operational estimates accessible, but their values remain model- and environment-dependent \cite{anthony2020carbontracker,codecarbon}. Henderson \emph{et al.} therefore recommend systematic reporting of hardware, runtime, energy, and carbon boundaries \cite{henderson2020systematic}. We follow that principle and treat carbon as one measured view of sustained performance, not as a hardware-independent constant or a full life-cycle assessment.

HPC energy efficiency has similarly moved from peak throughput toward application-level work per watt and cross-layer optimization \cite{feng2007green500}. The lesson is directly relevant to hybrid AI pipelines: an accelerator kernel cannot be evaluated independently of data movement, host orchestration, materialized features, and downstream processing. Our benchmark adopts an application boundary that includes reservoir feature generation, readout training, and inference, and it reports memory and latency alongside energy.

\subsection{Reservoir and Neural Forecasting}
Reservoir computing fixes a nonlinear recurrent system and trains only a readout \cite{jaeger2001echo,lukosevicius2012practical}. Echo State Networks (ESNs) are therefore a critical baseline for QRC: both expose a high-dimensional temporal state while avoiding end-to-end recurrent training. We also include persistence, ARIMA, lagged ridge regression, MLP, RNN, GRU, LSTM, TCN, and Transformer models. TCNs offer long receptive fields and parallel sequence processing \cite{bai2018empirical}; LSTM and GRU use gated recurrence \cite{hochreiter1997long,cho2014learning}; Transformers use attention-based sequence modeling \cite{vaswani2017attention}. Including cheap and strong baselines is necessary for an energy-efficiency claim: efficiency relative only to an unnecessarily expensive model is not meaningful.

\subsection{Quantum and Quantum-Inspired Temporal Models}
QRC exploits fixed quantum dynamics as a temporal feature map \cite{fujii2017harnessing}. Spatial multiplexing, continuous-variable Gaussian reservoirs, natural quantum dynamics, neutral atoms, and few-atom cavity systems broaden the design space \cite{nakajima2019boosting,nokkala2021gaussian,suzuki2022natural,kornjaca2024large,zhu2025practical}. Recent studies apply QRC to chaotic forecasting, finance, and industrial degradation \cite{ahmed2025robust,li2025volatility,tandon2025corrosion}. In contrast, variational QML trains parameterized circuits and may incur gradient, measurement, and barren-plateau costs \cite{mcclean2018barren,havlicek2019supervised}. These distinctions matter for energy accounting: fixed reservoirs shift cost toward feature generation and reuse, while variational circuits repeatedly invoke simulation or hardware during optimization.

Sustainable quantum computing ultimately requires life-cycle accounting for cryogenics, control electronics, fabrication, utilization, and end-of-life impacts \cite{arora2024sustainableqc}. Our scope is narrower and directly measurable: the operational cost of the classical/simulator workflow used in current QRC research. We make no claim that simulator energy predicts QPU energy.

\section{Benchmark and Measurement Methodology}
\subsection{Workloads and Model Matrix}
The benchmark uses five one-step nonlinear forecasting tasks: NARMA-10, NARMA-20, Mackey--Glass, Lorenz-63, and the Santa Fe laser series \cite{mackey1977oscillation,lorenz1963deterministic,weigend1994time}. Each model uses chronological 60/20/20 train/validation/test splits, training-only normalization, a fixed lookback (64 in the common protocol), horizon one, and seeds $\{0,1,2,3,4\}$. Fixed model configurations are used rather than hidden hyperparameter sweeps, isolating the measured execution cost of each configuration.

The 33 configurations comprise ten classical/non-QRC models, two variational QML models, and 21 QRC--readout combinations. The QRC encoders are: (i) \cvb, a continuous-variable Gaussian moment simulator with 8 modes and 4 virtual nodes; (ii) \cvh, a 16-mode, 8-virtual-node feature map; and (iii) \gbq, a 6-qubit, 2-virtual-node gate-based statevector simulator. Each encoder is paired with linear, MLP, RNN, GRU, LSTM, TCN, and Transformer readouts. The two additional QML models are a four-qubit QNN and an Ising-entangling QNN, each with two circuit layers and ten epochs. Table~\ref{tab:matrix} summarizes the system roles.

\begin{table}[t]
\centering
\caption{Benchmark matrix and dominant execution stage.}
\label{tab:matrix}
\scriptsize
\setlength{\tabcolsep}{3.2pt}
\begin{tabular}{p{1.35cm}p{2.55cm}p{3.15cm}}
\toprule
Family & Configurations & Dominant measured stage \\
\midrule
Statistical / linear & Persistence, ARIMA, ridge-lag & CPU fitting or closed-form solve \\
Classical reservoir & ESN & Reservoir rollout + linear readout \\
Neural sequence & MLP, RNN, GRU, LSTM, TCN, Transformer & GPU/CPU training and inference \\
CV-QRC & 2 encoders $\times$ 7 readouts & Moment-feature generation, storage, readout \\
GB-QRC & 1 encoder $\times$ 7 readouts & Statevector circuit evaluations, readout \\
Variational QML & QNN, QNN-Ising & Repeated circuit simulation in optimization \\
\bottomrule
\end{tabular}
\end{table}

\subsection{Application Boundary and Instrumentation}
Figure~\ref{fig:boundary} shows the measured application path. A run loads or generates a temporal series, constructs windows, optionally generates and materializes reservoir features, trains a readout, and performs test inference. The tracker records elapsed time, operational energy, carbon estimate, peak RAM, peak GPU memory, and per-sample inference latency. Each run writes one JSON record containing metrics, hardware identifier, backend type, model size, feature dimension, and cache metadata. The core artifact includes 775 fully instrumented records (31 configurations $\times$ 5 datasets $\times$ 5 seeds). The separate QNN and QNN-Ising sweep contributes 50 additional completed runs.

\begin{figure}[t]
\centering
\resizebox{\columnwidth}{!}{%
\begin{tikzpicture}[
  node distance=2.8mm,
  box/.style={draw,rounded corners,align=center,minimum height=6.5mm,text width=1.28cm,font=\scriptsize},
  wide/.style={draw,rounded corners,align=center,minimum height=6.5mm,text width=1.55cm,font=\scriptsize},
  arr/.style={-{Latex[length=1.6mm]},thick},
  group/.style={draw,dashed,rounded corners,inner sep=2mm}
]
\node[box] (data) {Series\\window};
\node[wide,right=of data] (enc) {Fixed QRC /\\classical encoder};
\node[wide,right=of enc] (cache) {Materialized\\feature cache};
\node[box,right=of cache] (head) {Readout\\training};
\node[box,right=of head] (infer) {Test\\inference};
\draw[arr] (data)--(enc);
\draw[arr] (enc)--(cache);
\draw[arr] (cache)--(head);
\draw[arr] (head)--(infer);
\node[group,fit=(enc)(cache),label={[font=\scriptsize]below:reusable stage}] {};
\node[group,fit=(head)(infer),label={[font=\scriptsize]below:per-readout stage}] {};
\end{tikzpicture}%
}
\caption{Measured application boundary. QRC feature generation and materialization are reusable; readout training and inference are readout-specific. Classical models bypass or internalize the reusable stage.}
\label{fig:boundary}
\end{figure}

Energy and carbon are estimated consistently on one local Windows workstation with an NVIDIA GTX 1650-class GPU. The measurements support controlled relative comparison on this platform. They do not include embodied emissions, facility PUE beyond the software estimator, networking, hardware manufacture, or QPU infrastructure. The fully instrumented 775-run core logged approximately 13.27 wall-clock hours including cache creation, 285 Wh, and 202 \gco{}. The 50 variational-QML runs add approximately 2.87 hours, 46.6 Wh, and 32.6 \gco{} based on their per-run means. These totals describe the executed benchmark, not cold-start recharging of reused caches.

\subsection{Cold-Start and Amortized Feature Accounting}
For dataset $d$, QRC encoder $q$, seed $s$, and readout $r$, let $C^{\mathrm{cache}}_{dqs}$ denote feature-generation carbon and $C^{\mathrm{head}}_{dqrs}$ the readout/inference carbon recorded after the cache exists. With $R=7$ readouts sharing one feature tensor, we report
\begin{align}
C^{\mathrm{cold}}_{dqrs} &= C^{\mathrm{head}}_{dqrs}+C^{\mathrm{cache}}_{dqs}, \\
C^{\mathrm{amort}}_{dqrs} &= C^{\mathrm{head}}_{dqrs}+C^{\mathrm{cache}}_{dqs}/R.
\end{align}
The same decomposition is applied to energy and wall time. Cold-start accounting is appropriate for a single readout or one-off deployment. Amortized accounting is appropriate when the feature tensor is intentionally shared across readouts, repeated probes, uncertainty heads, or related downstream tasks. Neither view is universally correct; both are necessary to expose the reuse assumption.

\subsection{Task and Sustainable-Performance Metrics}
Forecast quality is reported using RMSE, MAE, and standard-deviation-normalized RMSE (NRMSE). System metrics are final training time, feature-cache time, inference latency per sample, peak RAM, peak GPU memory, trainable and total parameters, energy, and carbon. We use NRMSE for cross-dataset comparisons.

A scalar diagnostic is useful for ranking, but it must not replace raw metrics. For each dataset $d$ and model $m$, we normalize skill between persistence and the best observed NRMSE:
\begin{equation}
A_{dm}=\mathrm{clip}_{[0,1]}\!\left(\frac{\mathrm{NRMSE}_{d,\mathrm{pers}}-\mathrm{NRMSE}_{dm}}{\mathrm{NRMSE}_{d,\mathrm{pers}}-\min_j\mathrm{NRMSE}_{dj}}\right).
\end{equation}
Carbon efficiency is min--max normalized in log space because costs span orders of magnitude:
\begin{equation}
E_{dm}=1-\frac{\log_{10}(C_{dm}+\epsilon)-\min_j\log_{10}(C_{dj}+\epsilon)}{\max_j\log_{10}(C_{dj}+\epsilon)-\min_j\log_{10}(C_{dj}+\epsilon)}.
\end{equation}
The Sustainable Forecasting Score is $\mathrm{SFS}_{dm}=\sqrt{A_{dm}E_{dm}}$. The geometric mean penalizes a model that fails either skill or efficiency. We report SFS under both accounting views and cross-check every conclusion against NRMSE, carbon, latency, memory, and the Pareto frontier.

\section{Results}
\subsection{Accuracy and Sustainable Performance Select Different Winners}
Table~\ref{tab:overall} reports representative configurations averaged across the five datasets. The best mean NRMSE is achieved by \cvb+Transformer (0.332), followed by \cvb+TCN (0.340), classical TCN (0.341), and classical Transformer (0.350). These differences are small compared with the resource spread. ESN and ridge-lag have worse average NRMSE (0.437 and 0.470) but require only 0.014 and 0.012 amortized \gco{} per run and obtain the highest SFS values (0.814 and 0.805). Thus, accuracy leaders are not sustained-efficiency leaders.

\begin{table*}[t]
\centering
\caption{Representative average results over five datasets. Carbon is \gco{} per run. QRC rows show amortized and cold-start values; classical rows are identical under both accounting views.}
\label{tab:overall}
\scriptsize
\setlength{\tabcolsep}{4.4pt}
\begin{tabular}{lrrrrrrrr}
\toprule
Model & NRMSE$\downarrow$ & SFS$_a$$\uparrow$ & SFS$_c$$\uparrow$ & CO$_{2,a}$$\downarrow$ & CO$_{2,c}$$\downarrow$ & Time$_a$ (s) & Lat.(ms) & Peak RAM(GB) \\
\midrule
ESN & 0.437 & \textbf{0.814} & \textbf{0.815} & 0.014 & 0.014 & 16.3 & 3.32 & 0.54 \\
Ridge-lag & 0.470 & 0.805 & 0.805 & \textbf{0.012} & \textbf{0.012} & 15.4 & 3.22 & 0.54 \\
MLP & 0.366 & 0.769 & 0.784 & 0.051 & 0.051 & 22.0 & 2.73 & 2.24 \\
\cvb+Linear & 0.407 & 0.690 & 0.559 & 0.021 & 0.098 & 10.3 & 2.65 & 2.69 \\
TCN & 0.341 & 0.602 & 0.639 & 0.186 & 0.186 & 38.4 & 3.17 & 2.92 \\
\cvb+TCN & 0.340 & 0.592 & 0.576 & 0.182 & 0.260 & 43.0 & 3.12 & 3.34 \\
\cvb+Transformer & \textbf{0.332} & 0.510 & 0.524 & 0.313 & 0.390 & 62.8 & 3.13 & 2.84 \\
Transformer & 0.350 & 0.510 & 0.569 & 0.338 & 0.338 & 57.7 & 3.41 & 2.33 \\
\cvh+TCN & 0.345 & 0.013 & 0.010 & 1.105 & 1.626 & 199.4 & 6.81 & 4.93 \\
\gbq+Linear & 1.003 & 0.000 & 0.000 & 0.128 & 0.848 & 43.4 & 2.62 & 2.34 \\
\bottomrule
\end{tabular}
\end{table*}

The dataset-level winners are heterogeneous. A small RNN is best on NARMA-10; \cvb+Transformer is best on NARMA-20; a classical Transformer is best on Mackey--Glass; and \cvb+TCN is best on Lorenz-63 and Santa Fe. This pattern supports a restrained conclusion: QRC can supply useful nonlinear temporal features, but its advantage is dataset- and readout-dependent rather than universal.

The two variational QML baselines do not alter the frontier. Across five datasets and five seeds, QNN obtains mean NRMSE 0.376, 189 s, and 0.573 \gco{} per run. QNN-Ising improves NRMSE to 0.367 but increases time to 224 s and carbon to 0.731 \gco{}. The Ising variant is useful on Lorenz-63, where its mean NRMSE improves from 0.088 to 0.060, but the modest aggregate gain does not offset simulator cost relative to efficient classical or fixed-reservoir models.

\subsection{The Pareto Frontier Favors Lightweight Baselines}
\begin{figure*}[t]
\centering
\includegraphics[width=0.82\textwidth]{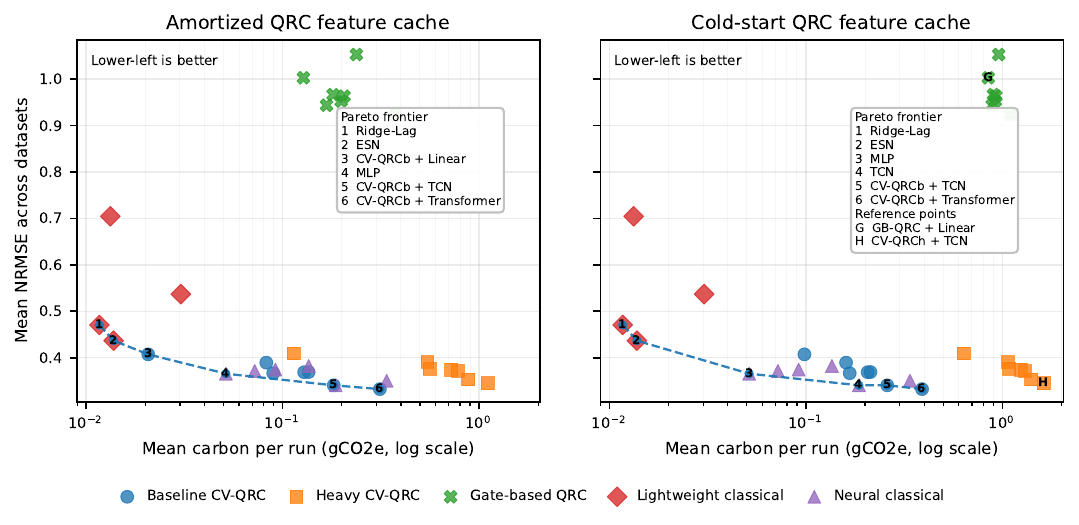}
\caption{Accuracy--carbon Pareto views averaged over datasets. The left panel amortizes one QRC feature cache over seven readouts; the right panel charges the full cache to every QRC configuration. Lower and farther left are better.}
\label{fig:pareto}
\end{figure*}

Figure~\ref{fig:pareto} shows average NRMSE versus carbon under amortized and cold-start accounting. ESN, ridge-lag, and MLP occupy the most useful low-carbon region. Classical TCN and \cvb+TCN are nearly indistinguishable in average NRMSE, but neither dominates lightweight methods in carbon. The QRC-linear configuration is notable: after amortization it combines moderate NRMSE with low readout cost and moves closer to the frontier. Under cold-start accounting, the same model shifts right because the one-time feature cost is visible.

This result is important for HPC/AI evaluation. A reusable accelerator stage can appear efficient or inefficient depending on the unit of accounting. Per-job accounting is appropriate when each job generates private features. Service-level accounting is appropriate when a feature cache is intentionally shared. The benchmark therefore treats the accounting boundary as a reported experimental parameter rather than a post-hoc choice.

\subsection{Stage-Level Cost Reveals the Value and Risk of Reuse}
Table~\ref{tab:family} aggregates the 775-run core by model family. The classical family consumes 33.2 Wh and 23.6 \gco{} across 250 runs. Baseline CV-QRC uses a comparable 33.4 Wh under amortized execution across 175 runs, but cold-start reassignment increases the hypothetical family total to 52.5 Wh. Heavy CV-QRC consumes 166.2 Wh as executed and 294.0 Wh under repeated cold starts. GB-QRC is the most accounting-sensitive: 52.3 Wh as executed versus 229.1 Wh if each readout regenerated its statevector features.

\begin{table}[t]
\centering
\caption{Core benchmark family totals. ``Executed'' reflects actual cache reuse in the 775 logs; ``cold'' reassigns the full cache to every QRC readout.}
\label{tab:family}
\scriptsize
\setlength{\tabcolsep}{3.4pt}
\begin{tabular}{lrrrr}
\toprule
Family & Runs & NRMSE & Wh$_{exec}$ & Wh$_{cold}$ \\
\midrule
Classical & 250 & 0.433 & 33.2 & 33.2 \\
\cvb & 175 & 0.368 & 33.4 & 52.5 \\
\cvh & 175 & 0.374 & 166.2 & 294.0 \\
\gbq & 175 & 0.972 & 52.3 & 229.1 \\
\bottomrule
\end{tabular}
\end{table}

The baseline CV reservoir illustrates productive reuse: one moderate-cost feature map supports readouts with different accuracy/cost points. The heavy and gate-based encoders illustrate the opposite risk: reuse can hide an oversized upstream stage if only per-readout cost is reported. A sustainable system should therefore expose both feature-service utilization and the marginal cost of each downstream head.

\subsection{Encoder--Readout Co-Design Matters}
Readout choice is not a minor implementation detail. For the same baseline CV encoder, a linear head has average NRMSE 0.407 and amortized carbon 0.021 \gco{}, whereas a Transformer reaches 0.332 but costs 0.313 \gco{}. TCN gives nearly the same accuracy as the Transformer at lower carbon, and its average NRMSE is essentially tied with the classical TCN. Recurrent heads are not consistently superior. This means QRC design is a system--application co-design problem: reservoir size, feature dimension, cache layout, readout architecture, and reuse count jointly determine sustained performance.

The larger CV reservoir is a negative scaling result. \cvh+TCN has worse mean NRMSE than \cvb+TCN (0.345 versus 0.340), while its amortized carbon is about six times larger (1.105 versus 0.182 \gco{}) and its peak RAM rises from 3.34 to 4.93 GB. Increasing modes, virtual nodes, and nonlinear feature products therefore fails the marginal-utility test. In accelerator terms, a larger intermediate representation increases feature generation, memory traffic, materialization, and readout work without improving the delivered application result.

\subsection{Statevector QRC Is Not a Sustainable Hardware Proxy}
GB-QRC obtains mean NRMSE near one and does not reach the useful frontier. On NARMA-10, its linear configuration performs roughly 995,000 circuit evaluations per seed yet yields error near persistence. Its feature generation is especially expensive under cold-start accounting. This is not evidence against hardware QRC; it is evidence that repeated dense statevector simulation is a poor proxy for sustainable hardware execution. Future studies should distinguish algorithmic QRC, simulator implementation, and physical backend rather than presenting ``quantum'' as a single energy category.

\begin{figure}[t]
\centering
\includegraphics[width=\columnwidth]{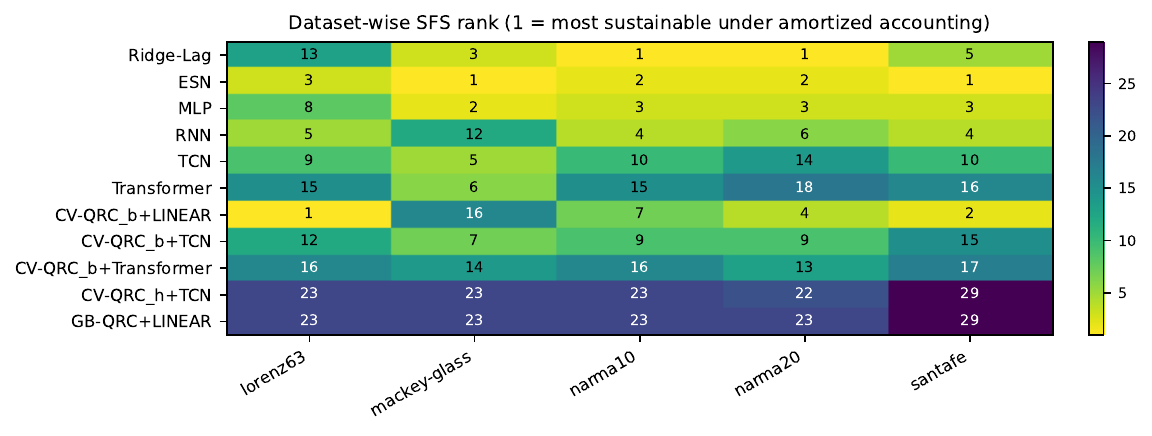}
\caption{Dataset-wise amortized SFS ranks for selected models. Lightweight models lead on NARMA and Mackey--Glass, while baseline CV-QRC variants are most competitive on Lorenz-63 and Santa Fe. Lower rank is better.}
\label{fig:ranks}
\end{figure}

\section{Implications for Sustainable HPC/AI Systems}
\subsection{Expose the Complete Application Path}
Energy claims should identify what is inside the measurement boundary. For a hybrid model, reporting only the trainable head omits feature generation; reporting only accelerator execution omits host work and materialization. At minimum, papers should separate data preparation, feature/embedding generation, model fitting, inference, and cache I/O. This decomposition also supports optimization: a readout-bound workload benefits from a more efficient head, while a feature-bound workload requires smaller representations, fewer simulator calls, or better reuse.

\subsection{Treat Reusable Features as a Scheduled Service}
A fixed reservoir resembles a shared embedding or feature service. When multiple consumers use the same tensor, the scheduler can batch generation, pin frequently used features, or place the cache near the accelerator that trains the heads. The relevant metric becomes useful downstream tasks per feature-generation joule. However, amortization must be justified by actual reuse, cache lifetime, storage, and invalidation. Our two accounting views define a practical bracket: cold start is a conservative upper bound; measured service-level reuse is the lower bound.

\subsection{Optimize for Marginal Utility, Not Representation Size}
The heavy CV result shows why larger feature spaces require explicit evidence. For every increase in modes, qubits, virtual nodes, shots, observables, or sequence length, the system should report marginal error reduction per additional joule and per additional byte. This is analogous to mixed precision, pruning, and model compression in conventional AI: the goal is not maximum internal dimension but maximum useful work under a budget.

\subsection{Use Pareto and Budgeted Views}
A single composite score is convenient but policy-dependent. SFS uses equal geometric weighting of skill and carbon efficiency; another deployment may prioritize latency, memory, or absolute energy. We therefore recommend publishing raw metrics and Pareto plots, followed by budgeted questions such as: What is the lowest NRMSE below 0.1 \gco{} per run? Which model fits within 1 GB RAM? How many readouts must share a cache before QRC becomes competitive? Such questions are actionable for Tier-1 systems, smaller research clusters, and local workstations alike.

\subsection{Separate Simulator Evidence from Hardware Claims}
The present benchmark quantifies simulator-based research cost. A QPU study must expand the boundary to include job queues, shots, calibration, control systems, cryogenics, error mitigation, and classical orchestration. Conversely, a hardware-native analog reservoir may avoid the exponential statevector cost observed here. A fair comparison should therefore report both the physical execution system and the classical workflow that enables it.

\section{Threats to Validity and Limitations}
The measurements come from one local machine and software-based energy/carbon estimation. Relative comparisons are controlled, but absolute values may change with hardware, utilization, operating system, power sensors, datacenter PUE, and grid intensity. The study is an operational benchmark, not a life-cycle assessment. Peak RAM and GPU memory are sampled rather than formally attributed to individual kernels.

The tasks are established nonlinear benchmarks rather than production finance, weather, or industrial workloads. They isolate memory and nonlinear dynamics but omit missing data, exogenous variables, distribution shift, probabilistic calibration, and long-horizon rollout. The common protocol uses one-step forecasting and fixed configurations; extensive hyperparameter optimization would alter both accuracy and total research cost. Our model matrix is broad but not exhaustive. QRNN and QLSTM seed sweeps were not completed because pilot simulations exceeded the available compute budget; we report completed QNN and QNN-Ising results instead and do not infer that recurrent QML is intrinsically inferior.

All QRC results are simulated. The CV moment model is an efficient quantum-inspired/simulated Gaussian reservoir, while GB-QRC uses statevectors. Neither measures QPU energy. Alternative tensor-network simulators, sparse circuit methods, analog hardware, finite-shot execution, and noise-aware encodings may change the frontier. Finally, SFS is a diagnostic, not a universal sustainability objective; raw metrics remain authoritative.

\section{Conclusion}
We presented a sustained-performance benchmark for nonlinear forecasting across classical, neural, variational-quantum, and simulated QRC models. The most accurate average configuration is baseline CV-QRC with a Transformer readout, but ESN and ridge-lag deliver the strongest energy--accuracy balance. Baseline CV-QRC can be competitive when features are reused, while heavy CV-QRC and gate-based statevector simulation consume substantially more energy without commensurate gains. The core lesson is systems-oriented: emerging AI workloads should be evaluated as complete pipelines, with stage-level energy, explicit cache accounting, memory and latency, and realistic reuse assumptions. Sustainable performance is not a property of a model label; it emerges from the co-design of representation, execution backend, data movement, readout, and scheduling.

\appendices
\section{Recommended Reporting Checklist}
For sustainable benchmarking of hybrid or accelerator-backed forecasting models, we recommend reporting: (1) the physical hardware or simulator backend; (2) host and accelerator identifiers; (3) model- and backend-specific resource parameters, such as qubits, modes, virtual nodes, shots, circuit evaluations, and feature dimension; (4) feature-generation, training, and inference time separately; (5) operational energy and carbon together with the measurement boundary; (6) cold-start and amortized costs when intermediate features are reused; (7) peak memory and inference latency; (8) inexpensive and strong classical baselines; and (9) raw measurements together with Pareto or budget-constrained performance views.

\section{Generative-AI Assistance Disclosure}
Generative-AI tools were used for language editing and assistance with code and LaTeX troubleshooting. They were not treated as sources of experimental evidence and are not credited as authors. The human authors designed and executed the experiments, inspected the raw logs, verified aggregate values and citations, selected the claims, and remain responsible for the manuscript and artifact.

\balance
\bibliographystyle{IEEEtran}
\bibliography{references}
\end{document}